\documentclass[aps,prb,showpacs,twocolumn,floats,psfig]{revtex4}
\usepackage{amssymb}
\usepackage{amsbsy}
\usepackage{amsmath}
\usepackage{epsfig}

\begin{document}

\date{\today}

\title {Josephson effect in graphene SBS junctions}

\author{Moitri Maiti and K. Sengupta }

\affiliation{TCMP division, Saha Institute of Nuclear Physics, 1/AF
Bidhannagar, Kolkata-700064, India. }

\date{\today}

\begin{abstract}

We study Josephson effect in graphene superconductor- barrier-
superconductor junctions with short and wide barriers of thickness
$d$ and width $L$, which can be created by applying a gate voltage
$V_0$ across the barrier region. We show that Josephson current in
such graphene junctions, in complete contrast to their conventional
counterparts, is an oscillatory function of both the barrier width
$d$ and the applied gate voltage $V_0$. We also demonstrate that in
the thin barrier limit, where $V_0 \rightarrow \infty$ and $d
\rightarrow 0$ keeping $V_0 d$ finite, such an oscillatory behavior
can be understood in terms of transmission resonance of
Dirac-Bogoliubov-de Gennes quasiparticles in superconducting
graphene. We discuss experimental relevance of our work.

\end{abstract}


\maketitle

\section{Introduction}

Graphene, a two-dimensional single layer of graphite, has been
recently fabricated by Novoselov {\it et.\,al.} \cite{nov1}.  In
graphene, the energy bands touch the Fermi energy at six discrete
points at the edges of the hexagonal Brillouin zone. Two of these
six Fermi points, referred to as $K$and $K'$ points, are
inequivalent and the quasiparticle excitations about them obey
linear Dirac-like energy dispersion \cite{ando1}. The presence of
such Dirac-like quasiparticles leads to a number of unusual
electronic properties in graphene including relativistic quantum
hall effect with unusual structure of Hall plateaus \cite{shar1},
which has been verified in experiments \cite{nov2}. Further, as
suggested in Ref.\ \onlinecite{geim1}, Dirac quasiparticles in
graphene leads to realization of interesting physical phenomenon
such as Klein paradox \cite{klein1,nov2}, Lorenz-boost type
phenomenon \cite{vinu1}, and unconventional Kondo effect
\cite{seng,g1}.

Another interesting consequences of the existence Dirac-like
quasiparticles can be understood by studying superconductivity in
graphene. It has been suggested that superconductivity can be
induced in a graphene layer in the presence of a superconducting
electrode near it via proximity effect
\cite{volkov1,beenakker1,beenakker2} or by intercalating it with
metallic atoms \cite{cneto1}. Consequently, studies on tunneling
conductance on both normal metal-superconductors (NS) and normal
metal-barrier-superconductor (NBS) junctions in graphene have been
undertaken \cite{beenakker1,sengupta1,sengupta2}. It has been shown
in Refs.\ \onlinecite{sengupta1} and \onlinecite{sengupta2} that the
tunneling conductance of such NBS junctions are oscillatory
functions of the effective barrier strength and that this
oscillatory phenomenon can be understood in terms of transmission
resonance phenomenon of Dirac-Bogoliubov-de Gennes (DBdG)
quasiparticles of graphene. Josephson effect has also been studied
in a superconductor-normal metal-superconducting (SNS) junction in
graphene \cite{beenakker2,ali1}. It has been shown in Ref.\
\onlinecite{beenakker2}, that the behavior of such junctions is
similar to conventional SNS junctions with disordered normal region.
Such Josephson junctions with thin barrier regions have also been
experimentally realized recently \cite{delft1}. However, Josephson
effect in graphene superconductor-barrier-superconductor (SBS)
junctions has not been studied so far.

In this work, we study Josephson effect in graphene for tunnel SBS
junctions. In this study, we shall concentrate on SBS junctions with
barrier thickness $d \ll \xi$ where $\xi$ is the superconducting
coherence length, and width $L$ which has an applied gate voltage
$V_0$ across the barrier region. Our central result is that in
complete contrast to the conventional Josephson tunnel junctions
studied so far \cite{likharev1,golubov1}, the Josephson current in
graphene SBS tunnel junctions is an oscillatory function of both the
barrier thickness $d$ and the applied gate voltage $V_0$. We provide
an analytical expression for the Josephson current of such a
junction. We also compute the critical current of graphene SBS
junctions. We find that this critical current is also an oscillatory
function of $V_0$ and $d$ and study the amplitude and periodicity of
its oscillation. We also show that in the thin barrier limit, where
the barrier region can be characterized by an effective
dimensionless barrier strength $\chi = V_0 d/\hbar v_F$ ($v_F$ being
the Fermi velocity of electrons in graphene), the Josephson current
becomes an oscillatory function of $\chi$ with period $\pi$
\cite{sengupta1}. We find that in this limit, the oscillatory
behavior of Josephson current can be understood as a consequence of
transmission resonance phenomenon of Dirac-Bogoliubov-de Gennes
(DBdG) quasiparticles in graphene. We demonstrate that the Josephson
current reaches the Kulik-Omelyanchuk limit \cite{ko1} for $\chi=n
\pi$ ($n$ being an integer), but, unlike conventional junctions,
never reaches the Ambegaokar-Baratoff limit \cite{ambe1} for large
$\chi$. We also discuss simple experiments to test our theory.

The organization of the rest of the paper is as follows. In Sec.\
\ref{se1}, we obtain an analytical expression for Josephson current
for a general SBS junction of thickness $d\ll \xi$ and applied
voltage $V_0$ and demonstrate that the Josephson current is an
oscillatory function of both $d$ and $V_0$. This is followed by
Sec.\ \ref{se2}, where we discuss the limiting case of thin barrier
and demonstrate that the oscillatory behavior of the Josephson
current can be understood in terms of transmission resonance of DBdG
quasiparticles in graphene. Finally we discuss experimental
relevance of our results in Sec.\ \ref{se3}.

\section{Josephson current for tunnel SBS junctions}
\label{se1}

\begin{figure}
\vspace{-1 cm} \rotatebox{0}{
\includegraphics[width=8cm]{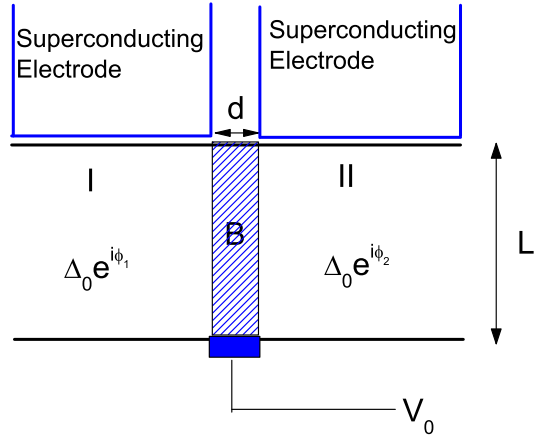}}
\caption{A schematic graphene SBS junction with the barrier B
sandwiched between two superconductors I and II with pair potentials
$\Delta_0 e^{i\phi_1}$ and $\Delta_0 e^{i\phi_2}$. The barrier
region is created by an external gate voltage $V_0$.} \label{fig1}
\end{figure}

We consider a SBS junction in a graphene sheet of width $L$ lying in
the $xy$ plane with the superconducting regions extending
$x=-\infty$ to $x=-d$ and from $x=0$ $x=\infty$ to for all $0\le y
\le L$, as shown in Fig.\ \ref{fig1}. The superconducting regions $x
\ge 0$ and $x \le -d$ shall be assumed to be kept close to
superconducting electrodes so that superconductivity is induced in
these regions via proximity effect \cite{volkov1,beenakker1}.
Alternatively, one can also possibly use intercalated graphene which
may have s-wave superconducting phases \cite{cneto1}. In the rest of
this work, we shall assume that these regions are superconducting
without worrying about the details of the mechanism used to induce
superconductivity. The region B, modeled by a barrier potential
$V_0$, extends from $x=-d$ to $x=0$. Such a local barrier can be
implemented by either using the electric field effect or local
chemical doping \cite{geim1,nov2}. In the rest of this work, we
shall assume that the barrier region has sharp edges on both sides
which requires $d \ll \lambda= 2\pi/k_F$, where $k_F$ and $\lambda$
are the Fermi wave-vector and Fermi wavelength for graphene. Such
barriers can be realistically created in experiments \cite{geim1}.
The width $L$ of the sample shall be assumed to be large compared to
all other length scales in the problem. The SBS junction can then be
described by the DBdG equations \cite{beenakker1}
\begin{eqnarray}
&&\left(\begin{array}{cc}
    {\mathcal H}_{a}-E_F + U({\bf r}) & \Delta ({\bf r}) \\
     \Delta^{\ast}({\bf r}) & E_F - U({\bf r})-{\mathcal H}_{a}
    \end{array}\right) \psi_{a}   = E \psi_{a}. \nonumber\\
\label{bdg1}
\end{eqnarray}
Here, $\psi_a = \left(\psi_{A\,a}, \psi_{B\,a}, \psi_{A\,{\bar
a}}^{\ast}, -\psi_{B\,{\bar a}}^{\ast}\right)$ are the $4$ component
wavefunctions for the electron and hole spinors, the index $a$
denote $K$ or $K'$ for electron/holes near $K$ and $K'$ points,
${\bar a}$ takes values $K'(K)$ for $a=K(K')$, $E_F$ denote the
Fermi energy, $A$ and $B$ denote the two inequivalent sites in the
hexagonal lattice of graphene, and the Hamiltonian ${\mathcal H}_a$
is given by
\begin{eqnarray}
{\mathcal H}_a &=& -i \hbar v_F \left(\sigma_x \partial_x + {\rm
sgn}(a) \sigma_y
\partial_y \right), \label{bdg2}
\end{eqnarray}
where ${\rm sgn}(a)$ takes values $\pm$ for $a=K(K')$.

The pair-potentials $\Delta({\bf r})$ in Eq.\ \ref{bdg1} connects
the electron and the hole spinors of opposite Dirac points. We have
modeled the pair-potential as
\begin{eqnarray}
\Delta({\bf r}) = \Delta_0 \left[\exp(i\phi_2) \theta(x) +
\exp(i\phi_1) \theta(x+d)\right], \label{pp}
\end{eqnarray}
where $\Delta_0$ is the amplitude and $\phi_{1(2)}$ are the phases
of the induced superconducting order parameters in regions I (II) as
shown in Fig.\ \ref{fig1}, and $\theta$ is the Heaviside step
function. Notice that the mean-field conditions for
superconductivity is satisfied as long as $\Delta_0\ll E_F$ or
equivalently $k_F \xi \gg 1$, where $\xi = \hbar v_F/\pi \Delta_{0}$
is the superconducting coherence length \cite{beenakker2}. The
potential $U({\bf r})$ gives the relative shift of Fermi energies in
the barrier and superconducting regions and is modeled as
\begin{eqnarray}
U({\bf r}) = V_0 \theta(-x) \theta(x+d). \label{poteq}
\end{eqnarray}

Solving Eq.\ \ref{bdg1}, we obtain the wavefunctions in the
superconducting and the barriers regions. In region I, for the DBdG
quasiparticles moving along $\pm x$ direction with a transverse
momentum $k_y=q = 2\pi n/L$ (for integer $n$) and energy $\epsilon$,
the wavefunctions are given by \cite{beenakker1}
\begin{eqnarray}
\psi_{I}^{\pm} &=& \left( u_1^{\pm}, u_2^{\pm},u_3^{\pm},u_4^{\pm}
\right) e^{\left[ i\left(\pm k_s x +q y\right) + \kappa x\right]},
 \label{supwave1}
\end{eqnarray}
where
\begin{eqnarray}
\frac{u_2^{\pm}}{u_1^{\pm}} &=& \pm \exp(\pm i\gamma), \quad
\frac{u_3^{\pm}}{u_1^{\pm}} = \exp[-i(\phi_1 \mp \beta)],\nonumber\\
\frac{u_4^{\pm}}{u_1^{\pm}} &=& \pm \exp[\pm i(\mp \phi_1 +\beta +
\gamma)], \label{ratieq1}
\end{eqnarray}
and $\sum_{i=1,4} |u_i|^2 \simeq 2\kappa$ is the normalization
condition for the wavefunction for $d \ll \kappa^{-1}$, where
$\kappa^{-1} = (\hbar v_F)^2 k_s/\left[E_F \Delta_0
\sin(\beta)\right]$ is the localization length. Here $k_s =
\sqrt{\left(E_F/\hbar v_F\right)^2 -q^2}$, $\gamma$, the angle of
incidence for the quasiparticles, is given by $\sin(\gamma) =\hbar
v_F q/E_F$, and $\beta$ is given by
\begin{eqnarray}
\beta  &=& \cos^{-1} \left(\epsilon/\Delta_0\right) \quad {\rm if}
\left|\epsilon\right| < \Delta_0 ,\nonumber\\
&=& -i \cosh^{-1} \left(\epsilon/\Delta_0\right) \quad {\rm if}
\left|\epsilon\right| > \Delta_0, \label{betaeq}
\end{eqnarray}
Note that for $\left|\epsilon\right| > \Delta_0$, $\kappa$ becomes
imaginary and the quasiparticles can propagate in the bulk of the
superconductor. The wavefunctions in region II ($x \ge 0$ ) can also
be obtained in a similar manner
\begin{eqnarray}
\psi_{II}^{\pm} &=& \left( v_1^{\pm}, v_2^{\pm},v_3^{\pm},v_4^{\pm}
\right) e^{\left[ i\left(\pm k_s x +q y\right) - \kappa x\right]},
 \label{supwave2}
\end{eqnarray}
where $\sum_{i=1,4} |v_i|^2=2\kappa$ and the coefficients $v_i$ are
given by
\begin{eqnarray}
\frac{v_2^{\pm}}{v_1^{\pm}} &=& \pm \exp(\pm i\gamma), \quad
\frac{v_3^{\pm}}{v_1^{\pm}} = \exp[-i(\phi_2 \pm \beta)],\nonumber\\
\frac{v_4^{\pm}}{v_1^{\pm}} &=& \pm \exp[\pm i(\mp \phi_1 -\beta +
\gamma)], \label{ratieq2}
\end{eqnarray}

The wavefunctions for electrons and holes moving along $\pm x$ in
the barrier region is given by
\begin{eqnarray}
\psi_B^{e \pm} &=& \left(1,\pm e^{\pm i \theta},0,0\right) \exp
\left[i\left(\pm k_{b} x + q y \right)\right]/\sqrt{2d}, \nonumber\\
\psi_B^{h \pm} &=&  \left(0,0,1,\pm e^{\mp i \theta'}\right) \exp
\left[i \left(\pm k'_{b} x + q y \right)\right]/\sqrt{2d}.
\label{barwave} \nonumber\\
\end{eqnarray}
Here the angle of incidence of the electron(hole) $\theta(\theta')$
is given by
\begin{eqnarray}
\sin\left[\theta(\theta')\right] &=& \frac{\hbar v_F q}{\epsilon
+(-)(E_F-V_0)} \nonumber\\
k_b (k'_b) &=& \sqrt{ \left(\frac{\epsilon +(-)(E_F-V_0)}{\hbar
v_F}\right)^2 -q^2} \label{bareq2}
\end{eqnarray}

To compute the Josephson current in the SBS junction, we now find
the energy dispersion of the subgap Andreev bound states which are
localized with localization length $\kappa^{-1}$ at the barrier
\cite{zagoskin1,kwon1}. The energy dispersion $\epsilon_n$
(corresponding to the subgap state characterized by the quantum
number $n$) of these states depends on the phase difference $\phi=
\phi_2-\phi_1$ between the superconductors. It is well known that
the Josephson current $I$ across the junction at a temperature $T_0$
is given by \cite{beenakker2,zagoskin1}
\begin{eqnarray}
I(\phi;\chi,T_0) &=& \frac{4e}{\hbar} \sum_{n} \sum_{q=-k_F}^{k_F}
\frac{\partial \epsilon_n}{\partial \phi} f(\epsilon_n),\label{jc1}
\end{eqnarray}
where $f(x)=1/(e^{x/(k_B T_0)}+1)$ is the Fermi distribution
function and $k_B$ is the Boltzman constant\cite{comment1}.

To obtain these subgap Andreev bound states, we now impose the
boundary conditions at the barrier. The wavefunctions in the
superconducting and barrier regions can be constructed using Eqs.\
\ref{supwave1}, \ref{supwave2} and \ref{barwave} as
\begin{eqnarray}
\Psi_I &=& a_1 \psi_I^{+}+ b_1 \psi_{I}^{-} \quad
\Psi_{II} = a_2 \psi_{II}^{+}+ b_2 \psi_{II}^{-}, \nonumber\\
\Psi_B &=& p \psi_B^{e +}+q \psi_B^{e -} + r \psi_B^{h +} + s
\psi_N^{h -}, \label{wave2}
\end{eqnarray}
where $a_1$($a_2$) and $b_1$($b_2$) are the amplitudes of right and
left moving DBdG quasiparticles in region I(II) and $p$($q$) and
$r$($s$) are the amplitudes of right(left) moving electron and holes
respectively in the barrier. These wavefunctions must satisfy the
boundary conditions:
\begin{eqnarray}
\Psi_I |_{x=-d} &=& \Psi_B |_{x=-d},  \quad  \Psi_{B} |_{x=0} =
\Psi_{II} |_{x=0}. \label{bc1}
\end{eqnarray}
Notice that these boundary conditions, in contrast their
counterparts in standard SBS interfaces \cite{kwon1}, do not impose
any constraint on derivative of the wavefunctions. Thus the standard
delta function potential approximation for short barriers
\cite{zagoskin1,kwon1} can not be taken the outset, but has to be
taken at the end of the calculations.

Substituting Eqs.\ \ref{supwave1}, \ref{supwave2}, \ref{barwave},
and \ref{wave2} in Eq.\ \ref{bc1}, we find the equations for
boundary conditions at $x=-d$ to be
\begin{eqnarray}
a_1 e^{-i k_s d -\kappa d} + b_1 e^{i k_s d -\kappa d} &=&
\nonumber\\
p e^{-i k_b d} + q e^{i k_b d} \nonumber\\
a_1 e^{i (\gamma-k_s d) -\kappa d} - b_1 e^{-i(\gamma- k_s d)
-\kappa d} &=& \nonumber\\
p e^{i(\theta-
k_b d)} - q e^{-i (\theta-k_b d)} \nonumber\\
a_1 e^{-i (\phi_1-\beta + k_s d) -\kappa d} + b_1
e^{-i(\phi_1+\beta-
k_s d) -\kappa d} &=& \nonumber\\
r e^{- ik'_b d} + s e^{i k'_b d} \nonumber\\
a_1 e^{i (\gamma-\phi_1 +\beta -k_s d) -\kappa d} - b_1
e^{-i(\gamma+\phi_1+\beta - k_s d) -\kappa d} &=&
\nonumber\\
r e^{-i(\theta'+ k_b d)} - s e^{i (\theta'+k'_b d)}. \label{bc2}
\end{eqnarray}
Similar equations at $x=0$ reads
\begin{eqnarray}
a_2 + b_2 &=& p+ q \nonumber\\
a_2 e^{i \gamma} - b_2 e^{- i \gamma} &=& p e^{i \theta} - q
e^{-i \theta} \nonumber\\
a_2 e^{-i(\phi_2 + \beta)} + b_2 e^{-i(\phi_2 -\beta)} &=& r+s
\nonumber\\
a_2 e^{i(\gamma -\phi_2 -\beta)} - b_2 e^{-i(\gamma+ \phi_2 -\beta)}
&=& re^{-i\theta'} - s e^{i\theta'} \label{bc3}
\end{eqnarray}
Eqs.\ \ref{bc2} and \ref{bc3} therefore yields eight linear
homogeneous equations for the coefficients $a_{i=1,2}$, $b_{i=1,2}$,
$p$, $q$, $r$, and $s$, so that the condition for non-zero solutions
of these coefficients can be obtained as
\begin{eqnarray}
{\mathcal A'} \sin(2 \beta) + {\mathcal B'} \cos(2 \beta) +
\mathcal{C'} =0 \label{bsd1}
\end{eqnarray}
where ${\mathcal A',\,B'},\,{\rm and}\,{\mathcal C'}$ are given by
\begin{eqnarray}
{\mathcal A'} &=& \cos(k'_b d) \cos(\gamma) \cos(\theta') \sin(k_b
d)
\left(\sin(\gamma) \sin(\theta)-1\right) \nonumber\\
&& + \cos(k_b d) \cos(\gamma) \cos(\theta) \sin(k'_b d)
\nonumber\\
&& + \frac{1}{2} \cos(k_b d) \cos(\theta) \sin(2\gamma)
\sin(\theta') \sin(k'_b
d)\nonumber\\
{\mathcal B'} &=& \sin(k'_b d) \sin(k_b d) \big [-1+
\sin(\theta)\sin(\gamma) \nonumber\\
&& -\sin(\theta')\sin(\gamma) + \sin(\theta)\sin(\theta')
\sin^2(\gamma) \big] \nonumber\\
&& -\cos(k_bd) \cos(k'_b d) \cos^2(\gamma) \cos(\theta)
\cos(\theta') \nonumber\\
{\mathcal C'} &=& \cos^2(\gamma) \cos(\theta) \cos(\theta')
\cos(\phi) - \sin(k_b d) \sin(k'_b d) \nonumber\\
&& \times \left[\sin(\theta) \sin(\theta')-\sin^2(\gamma) \right.
\nonumber\\
&& \left. + \sin(\gamma) \left( \sin(\theta) -\sin(\theta') \right)
\right] \label{cf1}
\end{eqnarray}
Note that in general the coefficients ${\mathcal A'}$, ${\mathcal
B'}$, and ${\mathcal C'}$ depends on $\epsilon$ through $k_b$,
$k'_b$, $\theta$ and $\theta'$ which makes it impossible to find an
analytical solution for Eq.\ \ref{bsd1}. However, for subgap states
in graphene SBS junctions, $\epsilon \le \Delta_0 \ll E_F$. Further,
for short tunnel barrier we have $|V_0 -E_F| \ge E_F$. In this
regime, as can be seen from Eqs.\ \ref{bareq2},  ${\mathcal A'}$,
${\mathcal B'}$, and ${\mathcal C'}$ become independent of
$\epsilon$ since $k_b \simeq  k'_b \simeq k_1 = \sqrt{[(E_F -
V_0)/\hbar v_F]^2 -q^2}$ and $\theta \simeq -\theta' \simeq \theta_1
= \sin^{-1}\left[\hbar v_F q/(E_F-V_0)\right]$ so that the
$\epsilon$ dependence of $k_b$, $k'_b$, $\theta$ and $\theta'$ can
be neglected. In this regime one finds that ${\mathcal A',B',C'}
\rightarrow {\mathcal A,B,C}$ where
\begin{figure}
\vspace{0 cm} \rotatebox{0}{
\includegraphics[width=\linewidth]{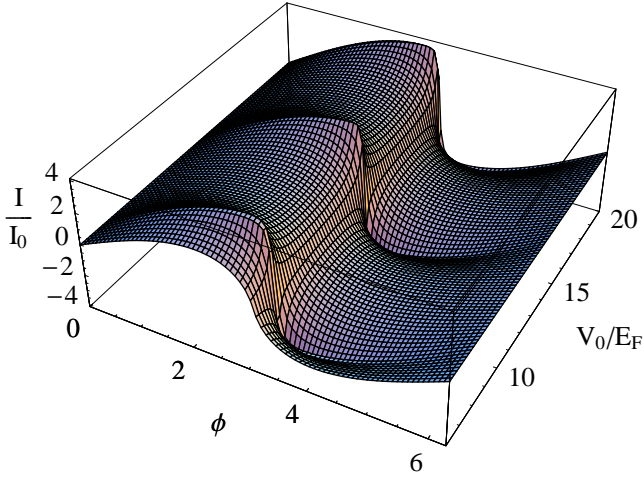}}
\caption{Plot of Josephson current $I$ as a function of phase
difference $\phi$ and the applied gate voltage $V_0$ for $k_B
T_0=0.01 \Delta_0$ and $d=0.5 \lambda$ showing oscillatory behavior
of $I/I_0$ as a function of the applied gate voltage.} \label{figm1}
\end{figure}
\begin{eqnarray}
{\mathcal A} &=& 0 \nonumber\\
{\mathcal B} &=& -\sin^2(k_1 d) \left[1-\sin(\gamma)\sin(\theta_1)
\right]^2 \nonumber\\
&& -\cos^2(k_1 d) \cos^2(\gamma) \cos^2(\theta_1) \nonumber\\
{\mathcal C} &=& \sin^2(k_1 d) \left[\sin(\gamma) - \sin(\theta_1)
\right]^2 \nonumber\\
&& + \cos^2(\gamma) \cos^2(\theta_1) \cos(\phi) \label{cf2}
\end{eqnarray}
The dispersion of the Andreev subgap states can now be obtained from
Eqs.\ \ref{bsd1} and \ref{betaeq}. One finds that there are two
Andreev subgap states with energies $\epsilon_{\pm} = \pm \epsilon$
where
\begin{eqnarray}
\epsilon &=& \Delta_0 \sqrt{1/2- {\mathcal C}/2{\mathcal B}}
\label{as1}
\end{eqnarray}
Using Eq.\ \ref{jc1}, one can now obtain the expression for the
Josephson current
\begin{eqnarray}
I(\phi,V_0,d,T_0) &=& I_0 g(\phi,V_0,d,T_0), \nonumber\\
g(\phi,V_0,d,T_0) &=& \int_{-\frac{\pi}{2}}^{\frac{\pi}{2}} d\gamma
\Bigg[ \frac{\cos^3(\gamma) \cos^2(\theta_1) \sin(\phi) }{{\mathcal
B}
\epsilon/\Delta_0} \nonumber\\
&& \times \tanh(\epsilon/2k_B T_0) \Bigg] \label{jc2a}
\end{eqnarray}
where $I_0 = e \Delta_0 E_F L/2\hbar^2 \pi v_F$ and we have replaced
$\sum_{q} \rightarrow E_F L/(2\pi \hbar v_F) \int_{-\pi/2}^{\pi/2} d
\gamma \cos(\gamma)$ as appropriate for wide junctions
\cite{beenakker2}.

\begin{figure}
\vspace{0 cm} \rotatebox{0}{
\includegraphics[width=\linewidth]{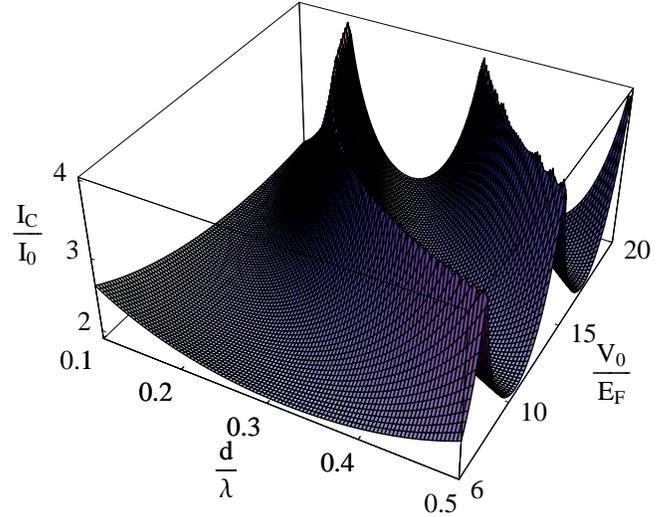}}
\caption{Plot of $I_c/I_0$ vs the applied gate voltage $V_0$ and the
junction thickness $d$ for $T_0=0.01 \Delta_0$.} \label{figm2}
\end{figure}

Eqs.\ \ref{as1}, and \ref{jc2a} represent the central result of this
work. From these equations, we find that both the dispersion of the
Andreev subgap states and the Josephson current in graphene SBS
junctions, in complete contrast to their conventional counterparts
\cite{likharev1,golubov1,zagoskin1}, is oscillatory function of the
applied gate voltage $V_0$ and the barrier thickness $d$. This
statement can be most easily checked by plotting the Josephson
current $I$ as a function of the phase difference $\phi$ and the
applied gate voltage $V_0$ for a representative barrier thickness
$d=0.5 \lambda$ and temperature $k_B T_0=0.01 \Delta_0$, as done in
Fig.\ \ref{figm1}. In Fig.\ \ref{figm2}, we plot the critical
current of these junctions $I_c(V_0,d,T_0)= {\rm
Max}[I(\phi,V_0,d,T_0)]$ as a function of the applied gate voltage
$V_0$ and barrier thickness $d$ for low temperature $k_B T_0 =0.01
\Delta_0$. We find that the critical current of these graphene SBS
junctions is an oscillatory function of both $V_0$ and $d$. This
behavior is to be contrasted with those of conventional junctions
where the critical current is a monotonically decreasing function of
both applied bias voltage $V_0$ and junction thickness $d$
\cite{likharev1,golubov1,zagoskin1}.

\begin{figure}
\vspace{0 cm} \rotatebox{0}{
\includegraphics[width=\linewidth]{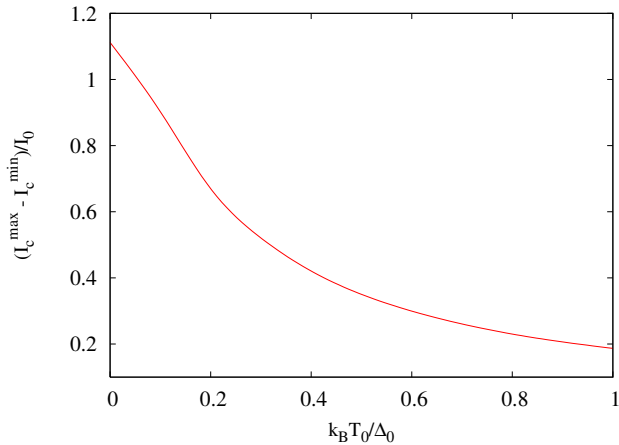}}
\caption{Plot of the temperature dependence of the amplitude of
oscillations of $I_c$ (given by $[I_c^{\rm max}(d)-I_c^{\rm
min}(d)]/I_0$) for $d=0.3 \lambda$. The amplitude is measured by
noting the maximum and minimum values of the critical current by
varying $V_0$ for a fixed $d$.} \label{figm3}
\end{figure}

Next, we analyze the temperature dependence of the amplitude of
oscillations of $I_c$. To find the amplitude of oscillation, we have
computed $I_c$ as a function of $V_0$ (for a representative value of
$d=0.3 \lambda$), noted the maximum ($I_c^{\rm max}$) and minimum
($I_c^{\rm min}$) values of $I_c$, and calculated the amplitude
$I_c^{\rm max}-I_c^{\rm min}$. The procedure is repeated for several
temperatures $T_0$ and the result is plotted in Fig.\ \ref{figm3}
which shows that the amplitude of oscillations decreases
monotonically as a function of temperature.

\begin{figure}
\vspace{0 cm} \rotatebox{0}{
\includegraphics[width=\linewidth]{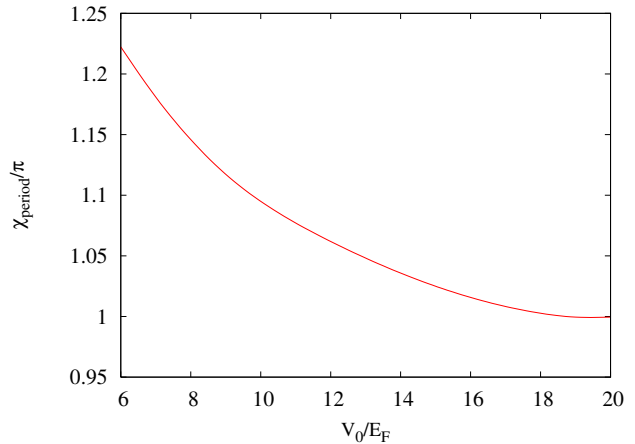}}
\caption{Plot of $\chi_{\rm period}$ of the critical current $I_c$
as a function of $V_0$. Note that $\chi_{\rm period}$ approaches
$\pi$ as we approach the thin barrier limit.} \label{figp1}
\end{figure}

Finally, we discuss the period of oscillation of the critical
current. To obtain the period, we obtain the critical current $I_c$
as a function of barrier width $d$ for the fixed applied gate
voltage $V_0$ and note down $d_{\rm period}$. We then compute
$\chi_{\rm period} = V_0 d_{\rm period}/\hbar v_F$ and plot
$\chi_{\rm period}$ as a function of $V_0$ for $k_B T_0= 0.01
\Delta_0$ as shown in Fig.\ \ref{figp1}. We find that $\chi_{\rm
period}$ decreases with $V_0$ and approaches an universal value
$\pi$ for large $V_0 \ge 20 E_F$. This property, as we shall see in
the next section, can be understood by analysis of graphene SBS
junctions in the thin barrier limit ($V_0 \rightarrow \infty$ and
$d\rightarrow 0$ such that $\chi= V_0 d /\hbar v_F$ remains finite
\cite{sengupta1}) and is a direct consequence of transmission
resonance phenomenon of DBdG quasiparticles in superconducting
graphene.

\section{Thin barrier limit}
\label{se2}

In the limit of thin barrier, where $V_0 \rightarrow \infty$ and
$d\rightarrow 0$ such that $\chi= V_0 d /\hbar v_F$ remains finite,
$\theta_1 \rightarrow 0$ and $k_1 d \rightarrow \chi$. From Eqs.\
\ref{cf2} and \ref{as1}, we find that in this limit, the dispersion
of the Andreev bound states becomes

\begin{eqnarray}
\epsilon_{\pm}^{\rm tb}(q,\phi;\chi)&=& \pm \Delta_0
\sqrt{1-T(\gamma,\chi)\sin^2(\phi/2)}, \label{abd}\\
T(\gamma,\chi) &=& \frac{\cos^2(\gamma)}{1-\cos^2(\chi)
\sin^2(\gamma)}. \label{teq}
\end{eqnarray}
where the superscript `tb' denote thin barrier limit. The Josephson
current $I$ can be obtained substituting Eq.\ \ref{teq} in Eq.\
\ref{jc1}. In the limit of wide junctions, one gets
\begin{eqnarray}
I^{\rm tb}(\phi,\chi,T_0) &=& I_0 g^{\rm tb} (\phi,\chi,T_0),
\nonumber\\
g^{\rm tb} (\phi,\chi,T_0)&=& \int_{-\pi/2}^{\pi/2} d \gamma \,
\Bigg[ \frac{T(\gamma,\chi) \cos(\gamma)
\sin(\phi)}{\sqrt{1-T(\gamma,\chi)\sin^2(\phi/2)}} \nonumber\\
&& \times \tanh \left(\epsilon_+ /2 k_B T_0\right) \Bigg].
\label{jc2}
\end{eqnarray}
Eqs.\ \ref{teq}, and \ref{jc2} represent the key results of this
section. From these equations, we find that the Josephson current in
graphene SBS junctions is a $\pi$ periodic oscillatory function of
the effective barrier strength $\chi$ in the thin barrier limit.
Further we observe that the transmission probability of the DBdG
quasiparticles in a thin SBS junction is given by $T(\gamma,\chi)$
which is also the transmission probability of a Dirac quasiparticle
through a square potential barrier as noted in Ref.\
\onlinecite{geim1}. Note that the transmission becomes unity for
normal incidence ($\gamma=0$) and when $\chi=n\pi$. The former
condition is a manifestation of the Klein paradox for DBdG
quasiparticles \cite{geim1}. However, this property is not reflected
in the Josephson current which receives contribution from
quasiparticles approaching the junction at all angles of incidence.
The latter condition ($\chi=n\pi$) represents transmission resonance
condition of the DBdG quasiparticles. Thus the barrier becomes
completely transparent to the approaching quasiparticles when
$\chi=n \pi$ and in this limit the Josephson current reduces to its
value for conventional tunnel junctions in the Kulik-Omelyanchuk
limit: $I^{\rm tb} (\phi,n\pi,T_0) = 4 I_0 \sin(\phi/2) {\rm
Sgn}(\cos(\phi/2)) \tanh \left(\Delta_0
\left|\cos(\phi/2)\right|/2k_B T_0\right)$ \cite{ko1}. This yields
the critical Josephson current $I_c^{\rm tb} (\chi=n\pi) = 4I_0$ for
$k_B T_0 \ll \Delta_0$. Note, however, that in contrast to
conventional junctions $T(\gamma,\chi)$ can not be made arbitrarily
small for all $\gamma$ by increasing $\chi$. Hence $I_c^{\rm tb}$
never reaches the Ambegaokar-Baratoff limit of conventional tunnel
junctions \cite{ambe1}. Instead, $I_c^{\rm tb}(\chi)$ becomes a
$\pi$ periodic oscillatory function of $\chi$. The amplitude of
these oscillations decreases monotonically with temperature as
discussed in Sec.\ \ref{se1}.

\begin{figure}
\vspace{0 cm} \rotatebox{0}{
\includegraphics[width=\linewidth]{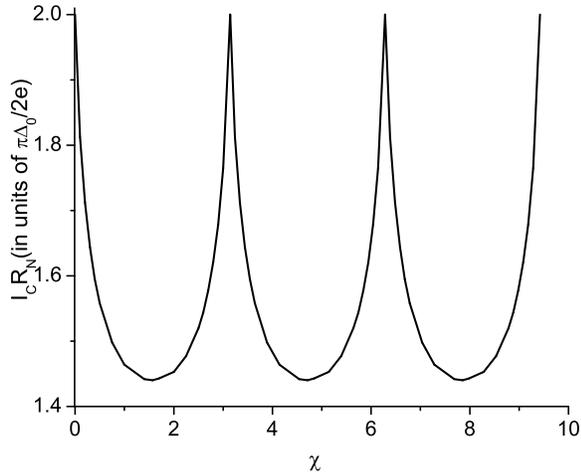}}
\caption{Plot of $I_c^{\rm tb} R_N$ as a function of $\chi$.
$I_c^{\rm tb} R_N$ is an oscillatory bounded function of $\chi$ and
never reaches its value ($\pi \Delta_0/2e$) for conventional
junctions in the Ambegaokar-Baratoff limit.} \label{figicrn}
\end{figure}

Finally, we compute the product $I_c^{\rm tb} R_N$ which is
routinely used to characterize Josephson junctions
\cite{likharev1,golubov1}, where $R_N$ is the normal state
resistance of the junction. For graphene SBS junctions $R_N$
corresponds to the resistance of a Dirac quasiparticle as it moves
across a normal metal-barrier-normal metal junction. For short and
wide junctions discussed here, it is given by $R_N = R_0/s_1(\chi)$
where $R_0  = \pi^2 v_F \hbar^2/(e^2 E_F L)$ and $s_1(\chi)$ is
given by \cite{geim1,beenakker2}
\begin{eqnarray}
s_1(\chi) &=& \int_{-\pi/2}^{\pi/2} d \gamma \, T(\gamma,\chi).
\cos(\gamma).
\end{eqnarray}
Note that $s_1(\chi)$ and hence $R_N$ is an oscillatory function of
$\chi$ with minimum $0.5 R_0$ at $\chi=n \pi$ and maximum $0.75 R_0$
at $\chi=(n+1/2)\pi$. The product $I_c^{\rm tb} R_N$, for thin SBS
junctions is given by
\begin{eqnarray}
I_c^{\rm tb} R_N &=& (\pi \Delta_0/2e) g^{\rm tb}_{\rm
max}(\chi,T)/s_1(\chi), \label{icrneq}
\end{eqnarray}
where $g_{\rm max}^{\rm tb} (\chi)$ is the maximum value of $g^{\rm
tb}(\phi,\chi)$. Note that $I_c^{\rm tb} R_N$ is independent of
$E_F$ and hence survives in the limit $E_F \rightarrow 0$
\cite{beenakker2}. For $k_B T_0 \ll \Delta_0$, $g_{\rm max}^{\rm tb}
(n\pi)=4$ and $s_1(n\pi)=2$, so that $I_c^{\rm tb} R_N|_{\chi=n\pi}
= \pi \Delta_0/e$ which coincides with Kulik-Omelyanchuk limit for
conventional tunnel junctions \cite{kwon1,ko1}. However, in contrast
to the conventional junction, $I_c^{\rm tb} R_N$ for graphene SBS
junctions do not monotonically decrease to the Ambegaokar-Baratoff
limit \cite{kwon1,ambe1} of $\pi \Delta_0/2e \simeq  1.57
\Delta_0/e$ as $\chi$ is increased, but demonstrates $\pi$ periodic
oscillatory behavior and remains bounded between the values $\pi
\Delta_0/e$ at $\chi=n\pi$ and $2.27 \Delta_0/e$ at
$\chi=(n+1/2)\pi$, as shown in Fig.\ \ref{figicrn}.

\section{Experiments}
\label{se3}

As a test of our predictions, we suggest measuring DC Josephson
current in these junctions as a function of the applied voltage
$V_0$. Such experiments for conventional Josephson junctions are
well-known \cite{ar1}. Further SNS junctions in graphene has also
been recently been experimentally created \cite{delft1}. For
experiments with graphene junctions which we suggest, the local
barrier can be fabricated by applying an additional gate voltage in
the normal region of the junctions studied in Ref.\
\onlinecite{delft1}. In graphene, typical Fermi energy can reach
$E_F \leq 80$meV with Fermi-wavelength $\lambda = 2\pi/k_F \geq
100$nm \cite{geim1}. Effective barrier strengths of $500-1000$meV
and barrier widths of $d \simeq 20-50$nm can be achieved in
realistic experiments \cite{geim1,nov2}. These junctions therefore
meet our theoretical criteria: $d \ll \lambda$ and $|V_0 -E_F| \ge
E_F$. To observe the oscillatory behavior of the Josephson current,
it would be necessary to change $V_0$ in small steps $\delta V_0$.
For barriers with fixed $d/\lambda=0.3$ and $V_0/E_F=10$, this would
require changing $V_0$ in steps of approximately $30$meV which is
experimentally feasible. The Joule heating in such junctions,
proportional to $I_c^2 R_N$, should also show measurable oscillatory
behavior as a function of $V_0$.

In conclusion, we have shown that the Josephson current in graphene
SBS junction shows novel oscillatory behavior as a function of the
applied bias voltage $V_0$ and the barrier thickness $d$. In the
thin barrier limit, such a behavior is the manifestation of
transmission resonance of DBdG quasiparticles in superconducting
graphene. We have suggested experiments to test our predictions.

KS thanks V. M. Yakovenko for discussions.

\end{document}